Continuous Measure of Symmetry as a Dynamic Variable: a New Glance on the Three-Body Problem

Mark Frenkel[a], Shraga Shoval,[b] Edward Bormashenko*[a]

[a]Department of Chemical Engineering, Ariel University, Ariel, POB 3, 407000, Israel

[b]Department of Industrial Engineering and Management, Faculty of Engineering, Ariel University, P.O.B. 3, Ariel 407000, Israel

*Correspondence to: Edward Bormashenko edward@ariel.ac.il

Abstract

The time evolution of the continuous measure symmetry for the system built of the three bodies interacting *via* the potential $U(r)\sim\frac{1}{r}$ is reported. Gravitational and electrostatic interactions between the point bodies were addressed. In the case of the pure gravitational interaction the three-body-system deviated from its initial symmetrical location, described by the Lagrange equilateral triangle, comes to collapse, accompanied by the growth of the continuous measure of symmetry. When three point bodies interact *via* the Coulomb repulsive interaction, the time evolution of CMS is quite different. CMS calculated for all of studied initial configurations of the point charges and all of their charge-to-mass ratios always comes with time to its asymptotic value, evidencing the stabilization of the shape of the triangle, constituted by the interacting bodies.

*Keywords*: three-body problem; Lagrange triangle; continuous measure of symmetry; gravity; Coulomb interaction; asymptotic value;

## 1. Introduction

In the three-body problem, three bodies/masses move in 3D space under their gravitational interactions as described by Newton's Law of gravity [1]. Solutions of this problem require that future and past motions of the bodies be uniquely determined based solely on their present positions and velocities. In general, the motions of the bodies take place in three dimensions (3D), and there are no restrictions on their masses nor on the initial conditions. This problem is referred as "the general three-body problem" [1-3]. Unlike two-body problems, no general closed-form solution of the three-body problem exists. Behavior of three-body dynamical systems is chaotic for most initial conditions, and numerical methods are generally required for deriving the trajectories of involved masses. In a restricted number of special configurations of the bodies, the exact solutions of the problem do exist. A special case of the three-body problem was analyzed by Euler [1-3]. Euler considered three bodies of arbitrary (finite) masses and placed them along a straight line. Euler demonstrated that the bodies would always stay on the same straight line for suitable initial conditions, and that the line would rotate about the center of mass of the system, resulting in periodic motions of all three bodies along ellipses [1-3]. Lagrange considered an equilateral triangle configuration of the three bodies, and demonstrated that in this case the bodies also move along elliptic orbits [1-3]. In the Lagrange solution, the initial configuration is an equilateral triangle and the three bodies are located at its vertices. We demonstrate that the Lagrange equilateral triangle enables the straightforward introducing of the continuous measure of symmetry to the analysis of the three-body problem. Lagrange proved that for suitable initial conditions, the triangular configuration is maintained and that the orbits of the three bodies remain elliptical for the duration of the motion [1-3]. The great progress in the development of the three-body problem is related to the fundamental studies by Henri Poincaré [4]. Poincaré and Bendixson studied conditions giving rise to existence of periodic solutions in the three-body problem [4].

Our paper presents the new approach to the three-body problem, based on the application of the continuous measure of symmetry (abbreviated CMS) to the problem. It seems that the first successful symmetrization of the equations of the three-body problem was performed in ref. 5. It was demonstrated, that the equations of the general three-body problem take on a very symmetric form when one considers only their relative positions, rather than position vectors relative to some given coordinate system [5]. From these equations one quickly derives some well-known classical properties of the three-body problem such as the first integrals and the equilateral triangle solutions [5]. We demonstrate the possibility to apply recently introduced CMS to the analysis of the three-body problem. The symmetry is usually considered within the binary paradigm; simply and roughly speaking, symmetry is present or absent in a given physical system. This YES/NO binary paradigm has been broken in refs. 6-11, in which the continuous measure of symmetry based on the symmetry distance of the shape, was introduced. The symmetry distance of the shape is defined as minimum mean square distance required to move points of the original shape in order to obtain a symmetrical shape [6-11]. Successful applications of CMS to analysis of physical problems is demonstrated in refs. 12-15.

## 2. Continuous measure of symmetry and its calculation

Let us acquaint the continuous symmetry measure, as it was defined in refs. 6-11. Consider a non-symmetrical shape consisting of $n_k$ points $M_i$, $(i = 1,2 \dots n_k)$ and a given symmetry group $G$. The continuous symmetry measure abbreviated CMS and denoted $S(G)$ is determined by the minimal average square displacement of the points $M_i$, that the shape has to undergo in order to acquire the prescribed $G$-symmetry. Assume that the $G$-symmetrical shape, emerges from the set of points $\widehat{M}_i$. Since the set $\widehat{M}_i$ is established, a CMS is defined as:

$$S(G) = \frac{1}{n_p \tilde{R}^2} \sum_{i=1}^{n_p} \left| M_i - \widehat{M}_i \right|^2, \quad (1)$$

where $\tilde{R}$ is the distance between the center of mass to the vertex of the most close equilateral triangle; (the squared values in Eq. 1 supply a function that is isotropic, continuous, and differentiable). The continuous measure of symmetry defined with Eq. 1 is a dimensionless value. At the first step, the points of the nearest shape possessing the $G$ - symmetry must be identified. An algorithm that identifies the set of points $\widehat{M}_i$ that constitute this symmetrical shape was introduced in refs. 6-11. **Figure 1** depicts equilateral triangle $M_{01}M_{02}M_{03}$ representing the symmetric shape that corresponds to the given non-symmetric triangle $M_1M_2M_3$.

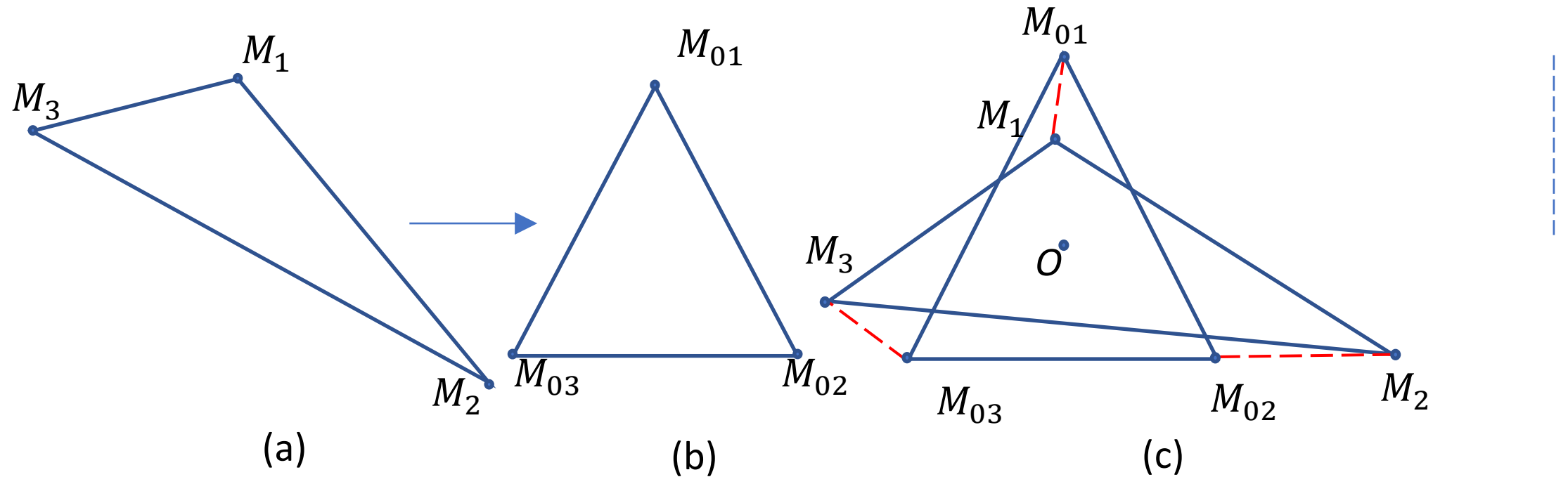

**Figure 1**. Given non-symmetric triangle $M_1M_2M_3$ (a). The equilateral triangle $M_{01}M_{02}M_{03}$ represents the symmetrical shape corresponding to the non-symmetric triangle $M_1M_2M_3$. Calculation of the CMS where point $O$ is the common centroid is shown (c).

The transformation of the non-symmetric triangle $M_1M_2M_3$ to the symmetric equilateral triangle $M_{01}M_{02}M_{03}$ is performed as follows: vertex $M_i$ is rotated counterclockwise around the common centroid $O$ of triangle $M_1M_2M_3$ by $\frac{2\pi(i-1)}{3}$ radians (one vertex of triangle $M_1M_2M_3$ remains fixed); thus, triangle $M_1M_2'M_3'$ emerges. Next, the location of the centroid $O'$of the intermediate triangle $M_1M_2'M_3'$ is determined. Centroid $O'$ is then rotated clockwise around the centroid $O$ by $-\frac{2\pi(i-1)}{3}$ radians (for the details see ref. 14).

Therefore, the equilateral triangle $M_{01}M_{02}M_{03}$ shown in **Figure 1**, represents the closest symmetrical shape to the pristine non-symmetrical triangle $M_1M_2M_3$ [6-14]. Since the set $\widehat{M}_i$ is established, the CSM is calculated with Eq. 1. Equilateral Lagrange triangle supplying the solution to the tree-body problem, hints to the effectivity of use of CMS for the solution of the three-body problem [1-2].

## 3. Symmetrized equations of motion for the three-body problem

Consider the three-body problem for set of three gravitating masses $m_i, i = 1 \ldots 3$. In the center-of-mass frame the equations of motion of the point gravitating masses appear as follows:

$$\ddot{x}_i = -Gm_j\frac{\vec{x}_i-\vec{x}_j}{|x_i-x_j|^3} - Gm_k\frac{\vec{x}_i-\vec{x}_k}{|x_i-x_k|^3} \quad (i,j,k = 1,2,3)\ , \qquad (2)$$

where $x_i, x_j\ , x_k(i,j,k = 1..3)$ are the coordinates of the masses in the center-of-mass frame defined by: $\sum_{i=1}^{i=3} m_i\vec{x}_i$=0; $\frac{d}{dt}\sum_{i=1}^{i=3} m_i\vec{x}_i$=0, illustrated with **Figure 2**, and G is the gravitational constant.

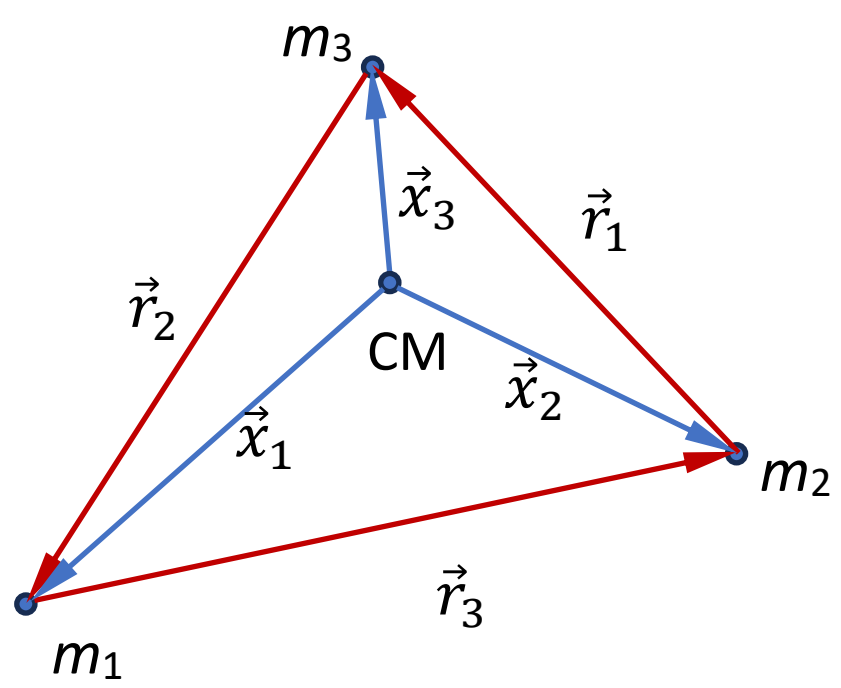

**Figure 2**. Location of gravitating masses $m_i, i = 1 \ldots 3$ in the center-of-mass frame, CM is the center of mass of the system; blue vectors depict $\vec{x}_i$ $(i = 1..3)$; red vectors depict the vectors of relative locations of the masses $\vec{r}_j$ $(j = 1..3)$ introduced in ref. 5.

Broucke and Lass in ref. 5 suggested the procedure of symmetrization of Eqs. 2, with the use of the vectors of relative locations of the masses, $\vec{r}_j = \vec{x}_m - \vec{x}_n(m = j-1, n =$ $j = n + 1$) (vector $\vec{r}_j$ corresponds to the side opposite the apex of the triangle occupied by mass $m_j$, see **Figure 2**). Eq. 3 is obviously true for vectors $\vec{r}_j$ ($j = 1..3$):

$$\sum_{j=1}^{3} \vec{r}_j = 0 \qquad (3)$$

Introducing vectors $\vec{r}_j, j = 1..3$ yields the equations of motion re-shaped and symmetrized as follows:

$$\frac{d^2}{dt^2}\vec{r}_i = -G\left(M\frac{\vec{r}_i}{r_i^3} - m_i\vec{R}\right), \qquad (4)$$

where $M = \sum_{i=1}^{3} m_i$ is the total mass of the system and vector $\vec{R}$ is defined with Eq. 5:

$$\vec{R} = \sum_{i=1}^{i=3} \frac{\vec{r}_i}{r_i^3} \qquad (5)$$

The first term in Eq. 4 is identical to that appearing on the standard two-body Kepler problem, whereas the second term in Eq. 4 generates the complexity of the problem. The Lagrange solution of the problem corresponds to the case when $r_1 = r_2 = r_3$ takes place. In this situation $\vec{R} = 0$ is true. Thus, the three-body problem is reduced to the two-body one and gravitating masses remain in the vertices of an equilateral triangle. The triangle may change its size and rotate; gravitating masses are moving along ellipses with different eccentricities; however, oriented by different angles one to another [16]. The motion of the gravitating masses in this case is periodic with the same period for all of the masses. It should be emphasized, that the aforementioned Lagrange solution remains stable only if one of the masses is much larger than other two ones [17-18].

## 4. Extension of the problem to the Coulomb interaction

Consider the system of tree point charges $q_i, i = 1,2,3$ possessing the corresponding masses $m_i, i = 1,2,3$. For a sake of simplicity, we assume that the motion of the point charges is slow; thus, the electrodynamic interaction between the charges may be neglected and the electrostatic and gravitational interactions between the bodies are essential. We consider the case when the initial velocities of the interacting bodies are zero; thus, our approach will be true at least at the initial stage of the motion, when the velocities of the bodies are still small (very roughly speaking, it is true when $\frac{v}{c} \ll 1$ is true, where $v$ is the velocity of the charge and $c$ is the light velocity in vacuum). The scalar equations of motion in this non-relativistic case appear as follows:

$$\ddot{x}_i = -(Gm_j - \frac{K}{m_i}q_iq_j)\frac{x_i - x_j}{|x_i - x_j|^3} - (Gm_k - \frac{K}{m_i}q_iq_k)\frac{x_i - x_k}{|x_i - x_k|^3} \quad , \qquad (6)$$

where K is the Coulomb constant. To make the problem even more simple, we also assume: $Gm_j \ll \frac{K}{m_i}q_iq_j$ and the gravitational interaction is negligible; this simple case yields interesting and understandable results, Thus, the equation of motion is re-written as follows:

$$\ddot{x}_i = \frac{K}{m_i}q_iq_j\frac{x_i - x_j}{|x_i - x_j|^3} + \frac{K}{m_i}q_iq_k\frac{x_i - x_k}{|x_i - x_k|^3} \qquad (7)$$

Symmetrical coordinates introduced in ref. 5 and discussed in detail in the previous Section yield in turn Eq. 8:

$$\frac{d^2}{dt^2}\vec{r}_i = K\left[\left(\frac{q_i}{m_i}q_j + \frac{q_j}{m_j}q_i + \frac{q_k}{m_k}q_k\right)\frac{\vec{r}_i}{r_i^3} - q_i\left(\frac{q_i}{m_i}\frac{\vec{r}_j}{r_j^3} + \frac{q_j}{m_j}\frac{\vec{r}_i}{r_i^3} + \frac{q_k}{m_k}\frac{\vec{r}_k}{r_j^3}\right)\right], i,j,k = 1 \ldots 3,$$

(8)

In the case when ${}^{q_1}/_{m_1} = {}^{q_2}/_{m_2} = {}^{q_3}/_{m_3} = \mu$ takes place, we obtain from Eq. 8, Eq. 9 which resembles Eq. 4, namely:

$$\frac{d^2}{dt^2}\vec{r}_i = K\mu\left(Q\frac{\vec{r}_i}{r_i^3} - q_i\vec{R}\right), \qquad (9)$$

where $Q = \sum_{i=1}^{3} q_i$ is the total electrical charge of the system and vector $\vec{R}$ is defined with Eq. 5. Eq. 9 immediately leads to the conclusion that the solution of the three-body problem similar to that suggested by Lagrange exists in the case when the interaction between the bodies is the electrostatic/Coulomb one. The Lagrange triangle appears as a solution of the three-body problem when the electrical charges are initially placed in the vertices of the equilateral triangle, and the condition ${}^{q_1}/_{m_1} = {}^{q_2}/_{m_2} = {}^{q_3}/_{m_3} = \mu = const$ takes place,

## 5. Three-body problem and the continuous measure of symmetry

The three-body problem in its general case has no analytical solution. We studied with the computer simulations the evolution of the continuous measure of symmetry of the three-body systems interacting *via* the Newtonian-Coulomb potential $U(r) \sim \frac{1}{r}$; the initial location of point bodies/charges was slightly shifted from their initial configuration, constituting the equilateral triangle (the so-called Lagrange triangle) .

Computer simulations were carried out with the software "Taylor Center" version 42; http://taylorcenter.org/Gofen/TaylorMethod.htm

Gravitational and Coulomb interactions were considered. In the first series of computer experiments the displacement of bodies from their initial location was 5%, as shown in **Figure 3**. In the second series of the numerical experiments the mass to charge ratio of the interacting particles was varied. In the third series of simulations the location of the particles and their charge-to-mass ratio were varied simultaneously.

We start from the first series of computer experiments, in which pure gravitational interaction between bodies is assumed. The displacement of the bodies from their initial location is illustrated with **Figure 3**; we shift the bodies along the coordinate axes; $\delta_1 = 0.05$ corresponds to the displacement of the mass $m_1$ to the right (in the positive direction of axis $x$), whereas $\delta_2 = 0.05$ corresponds to the displacement of the mass $m_1$ to the left (in the negative direction of axis $x$).

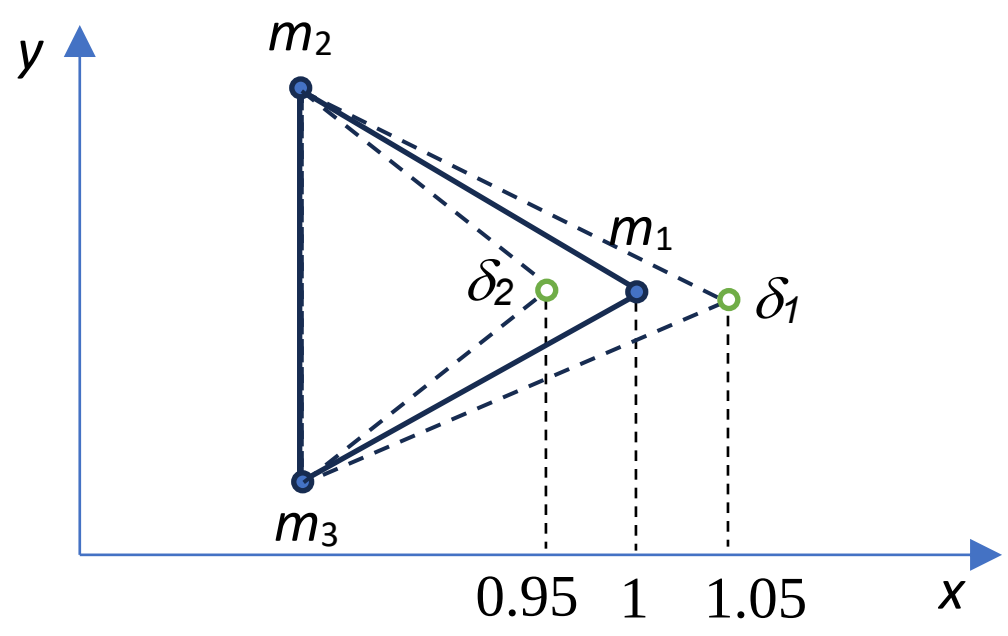

**Figure 3**. Deformation of the Lagrange triangle adopted in the numerical simulations is depicted. Point masses $m_i, i = 1 \dots 3$ are initially located in the vertices of the equilateral triangle. Diplacement $\delta_1 = 0.05$ corresponds to the displacement of the mass $m_1$ to the right (in the positive direction of axis $x$), displacement $\delta_2 = -0.05$ corresponds to the displacement of the mass $m_1$ to the left (in the negative direction of axis $x$).

Simple combinatory analysis demonstrates that there exists in total 64 possibilities of the distortion of the initial equilateral triangle, when locations of all of the vertices are perturbed; we do not report here the exhaustive analysis of all of the possible deformations of the Lagrange triangle, being focused on the some of illustrative examples of the general three-body problem. In all of the cases the continuous measure of symmetry (see Section 2) was taken as a dynamic variable $S(t)$ quantifying the distortion of the initial equilateral Lagrange triangle under the motion governed by gravitational or Coulomb forces (see Eq. 4 and Eq. 9). Consider the Example #1, in which three bodies interact by the pure gravitational interaction, and $x_1 = 1.05$ is assumed, as shown in **Figure 3** ($G = 1$ is assumed for a sake of simplicity). The initial positions of the gravitating bodies are: $m_1$(1.05, 0), $m_2$(cos(π/3), sin(π/3)), $m_3$(cos(2π/3), sin(2π/3)); the initial velocities are zero. The motion of two sets of dimensionless masses was analyzed, namely: $m_1 = m_2 = m_3 = 1$ and $m_1=3$, $m_2=4$, $m_3=5$. The results of calculations for both sets of masses are supplied in **Figure 4**.

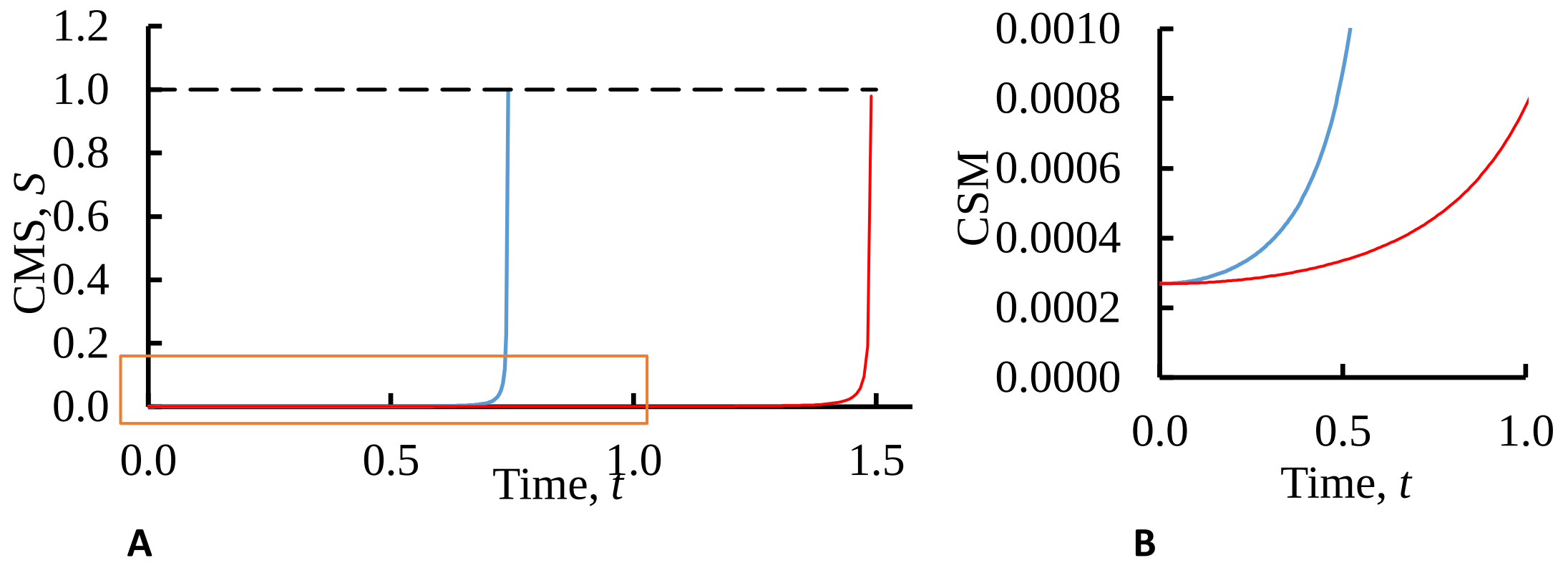

**Figure 4**. Time evolution of the continuous measure of symmetry (CMS) for a pure gravitational interaction of masses initially located in the vertices of the equilateral Lagrange triangle. Initial distortion corresponding to $x = 1.05$ is adopted. **A**. Inset (A) depicts the time evolution of CMS until the gravitational collapse of the tree-body gravitating system; **B**. Inset (B) demonstrates the initial stage of the motion. Initial velocities are zero. Blue curve corresponds to the set of masses $m_1=m_2=m_3=1$; red curve corresponds to the set of masses $m_1=3$, $m_2=4$, $m_3=5$.

**Figure 4** supplies a number of very important qualitative conclusions: i) irrespectively of the interrelation between the gravitating masses the distortion of the Lagrange triangle destroys the symmetry of the systems and results in the gravitational collapse of the system; the situation $CMS = 1$ corresponds to the disappearance of the one of the sides of the triangle, as it will be shown below; ii) destruction of the symmetry is not immediate; systems demonstrate certain stability until some threshold value of the distortion. This value needs additional physical and computational insights, which are not covered in the present paper.

## 6. Coulomb interaction, three-body problem and the continuous measure of symmetry

Now we consider the slightly deformed Lagrange triangle ($x = 1.05$ and $x = 0.95$ is true for one of the vertices). Point charges $q_i, i = 1..3$ are located in the vertices of the triangle, the initial velocities of the charges are zero. We address the situation, when electrodynamic interactions are neglected, and the Coulomb interactions dominate over gravitational ones (the Coulomb constant is equaled to unity). The masses of the bodies and their charges are: $m_1 = m_2 = m_3 = 1$; $q_1 = q_2 = q_3 = 1$. The equations of motion are supplied by Eq. 9. The results of the computer simulations, which establish the time evolution of CMS $S(t)$ are supplied in **Figure 5**.

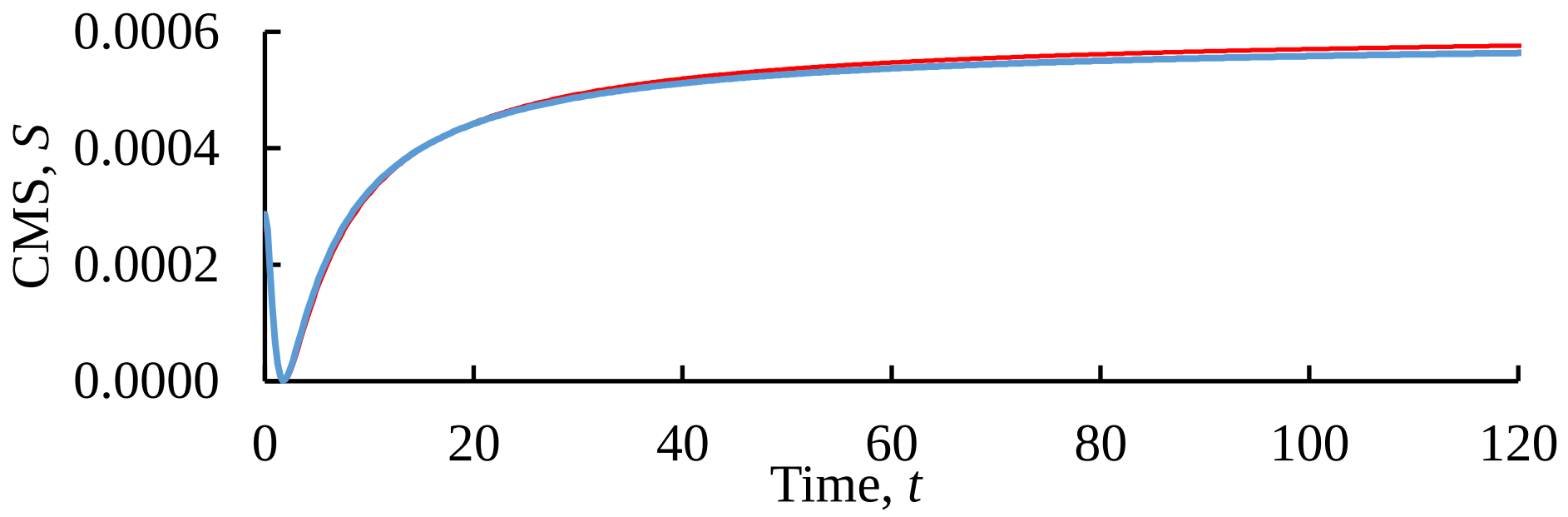

**Figure 5**. Time evolution of the continuous measure of symmetry calculated for the system of the identical point charges, located in the vertices of the distorted Lagrange triangle. Red line depicts the time evolution of CMS for $x$=1.05; the blue line shows the time evolution of CMS for $x$=0.95. Initial velocities of the point charges are zero; the masses are equal.

The time evolution of CMS demonstrates in both of the cases an identical behavior, which may be described as follows: the systems start with a non-zero value of CMS (the initial Lagrange triangle is distorted; afterwards, the charges came to the vertices of the equilateral triangle (CMS equals zero, see **Figure 5**), and afterwards the CMS grew and attained the asymptotic value. This result is intuitively clear: consider, that only repulsion Coulomb forces act between the point charges, these forces are weakened in a course of the motion of the charges; thus, the shape of the triangle constituted by the charges is stabilized, and consequently CMS attains its asymptotic value. This non-obvious result is of a primary importance, enabling qualitative characterization of the configuration of the moving point charges. Charges come to the vertices of the equilateral triangle with non-zero velocities, and thus, they pass these points and continue to move, driven by the repulsion Coulomb forces, finally obtaining the configuration, quantified by the asymptotic value of the CMS. We'll demonstrate that the asymptotic behavior of CMS is observed for various interrelations between the charges and masses of the interacting bodies.

Consider now the situation, when $m_1 >> m_2 = m_3$ and $q_1 >> q_2 = q_3$ takes place. These conditions are similar to those, inherent for the stable Lagrange solution of the three-body problem [1-2]. The time evolution of the continuous measure of symmetry for this case is depicted in **Figure 6**.

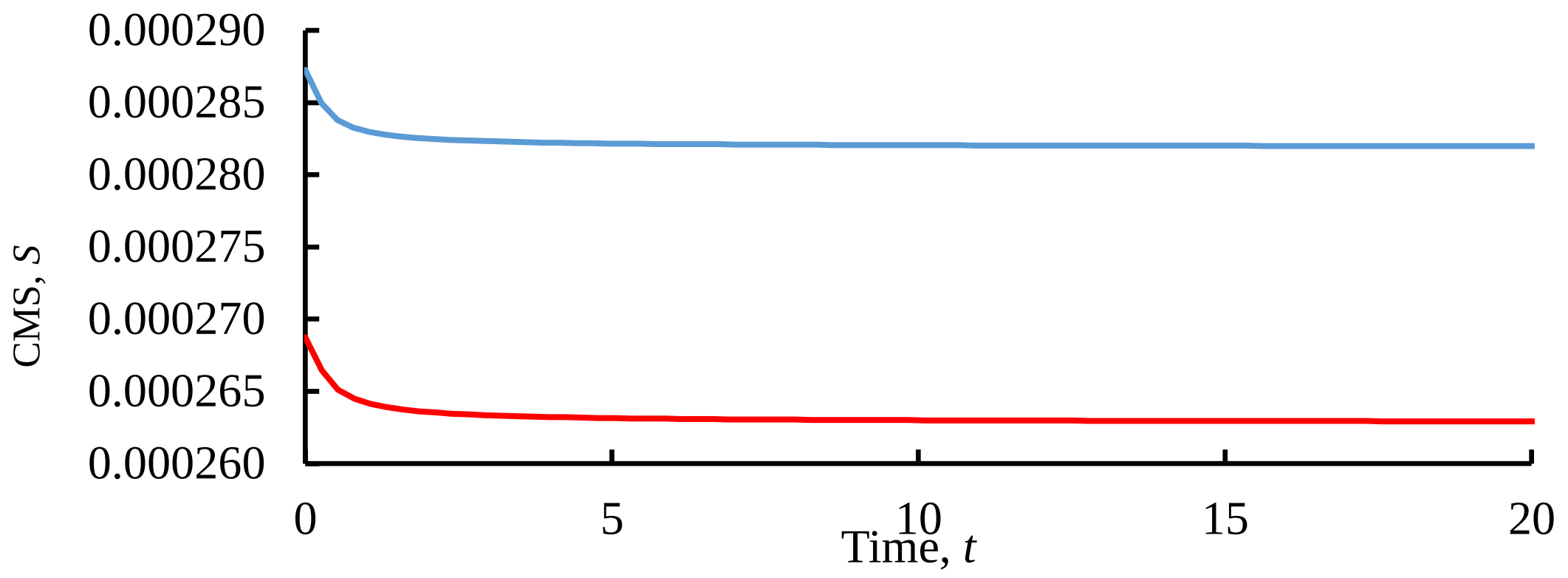

**Figure 6**. Time evolution of the continuous measure of symmetry calculated for the system of the point charges, located in the vertices of the distorted Lagrange triangle. $m_1$=1000; $m_2$= $m_3$ =1; $q_1$=1000; $q_2$=$q_3$=1 is assumed. Red line depicts the time evolution of CMS for $x$=1.05; the blue line shows the time evolution of CMS for $x$=0.95. Initial velocities of the point charges are zero

In this situation, the initial CMS is decreased in a course of the motion of the electrically charged bodies and it comes to its saturation value depending on the initial distortion of the Lagrange triangle, as shown in **Figure 6**. Again, only repulsive Coulomb interactions are present in the system; the Coulomb repulsion is decreased with time and eventually CMS attains its saturation value, quantifying the distortion of the initial triangle.

We also varied in our computer experiments the charge of one of the point masses ($q_1 = 0.95, q_1 = 1.05$ were tested numerically). In these experiments the charges were placed in the vertices of the undistorted equilateral Lagrange triangle. The time evolution of CMS is shown in **Figure 7**.

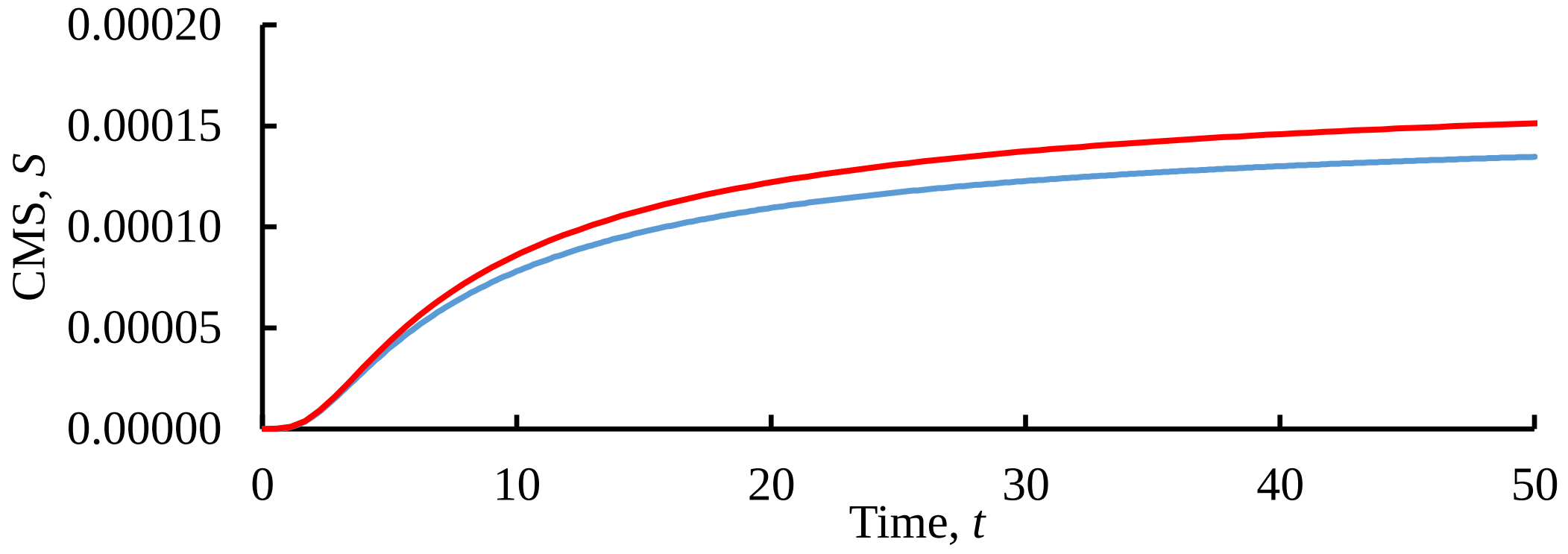

**Figure 7**. Time evolution of the continuous measure of symmetry calculated for the system of the point charges, located in the vertices of the non-distorted equilateral Lagrange triangle. $m_1 = m_2 = m_3 = 1$ is assumed; red line corresponds to $q_1$= 1.05; blue line corresponds to $q_1$=0.95. Initial velocities of the point charges are zero.

In this case, the initial value of the CMS is zero (the initial Lagrange triangle is equilateral). The value of CMS grows in a course of motion of the charges and comes to saturation, as shown in **Figure 7**.

We also tested the situation when the initial Lagrange triangle was slightly distorted and the charge-to-mass ratio of one of the charges was also slightly different from that prescribed for the other charges. The time evolution of CMS in this case is shown in **Figure 8**.

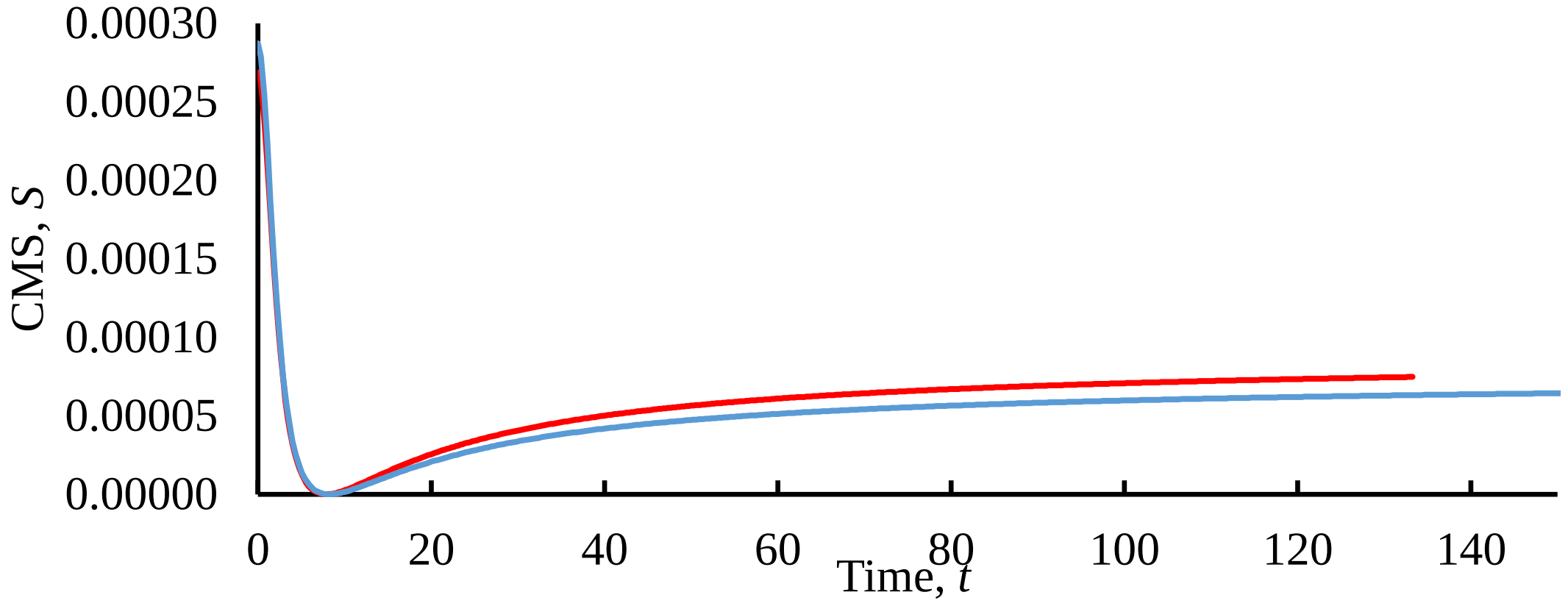

**Figure 8**. Time evolution of the continuous measure of symmetry calculated for the system of the point charges, located in the vertices of the distorted Lagrange triangle. $m_1$=$m_2$=$m_3$ =1; $q_1$=0.95/1.05; $q_2$ = $q_3$ =1 is assumed. Red line depicts the time evolution of CMS $S(t)$ for $x_1 = 1.05, q_1 = 1.05$; the blue line shows the time evolution of CMS for $x_1 = 0.95, q_1 = 0.95$. Initial velocities of the point charges are zero.

The time evolution of CMS in this case is similar to that shown in **Figure 5**, namely: the systems start with non-zero value of CMS (the initial Lagrange triangle is distorted); afterwards, the charges came to the vertices of the equilateral triangle with non-zero velocities (CMS equals zero, see **Figure 8**), and afterwards the CMS grows with time and attains its asymptotic value.

Now we perturb the geometrical symmetry and the charge-to-mass ratio for different vertices of the initial Lagrange triangle, namely we assume: $x_1$= 1.05 and $q_2$= 0.95 and $x_1$=0.95 and $q_2$=1.05. The temporal evolution of CMS is illustrated with **Figure 9**.

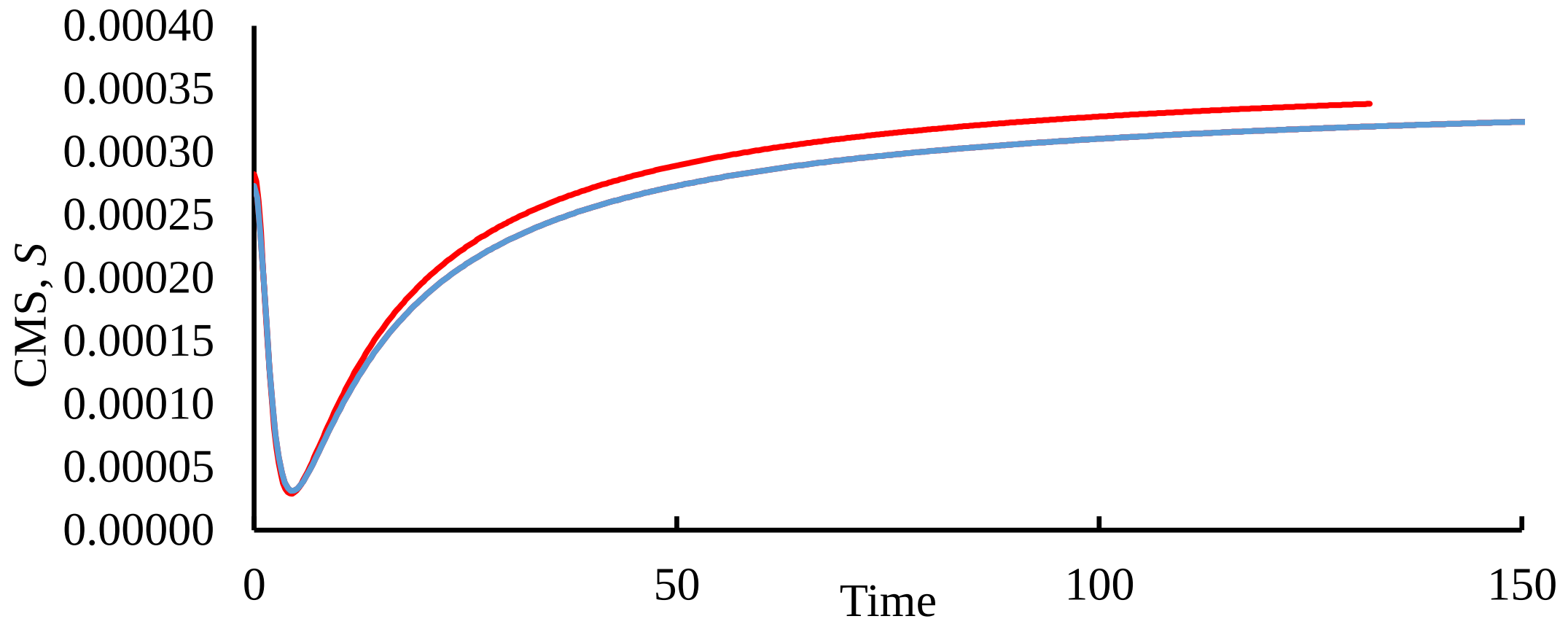

**Figure 9**. Time evolution of the continuous measure of symmetry calculated for the system of the point charges, located in the vertices of the distorted Lagrange triangle. $m_1$=$m_2$=$m_3$ =1; $q_1$=$q_3$ =1 is assumed. Red line corresponds to $x_1 = 1.05, q_2 = 0.95$; blue line corresponds to $x_1 = 0.95, q_1 = 1.05$. Initial velocities of the point charges are zero.

The time evolution of CMS resembles qualitatively the behavior of CMS depicted in **Figure 8**; however, the CMS does not attain zero as its minimal value, as it recognized from **Figure 9**.

The qualitative description of the change in the shape of the initial Lagrange triangle is illustrated with **Figure 10**, which supplies the main types of the time evolution of the Lagrange triangle.

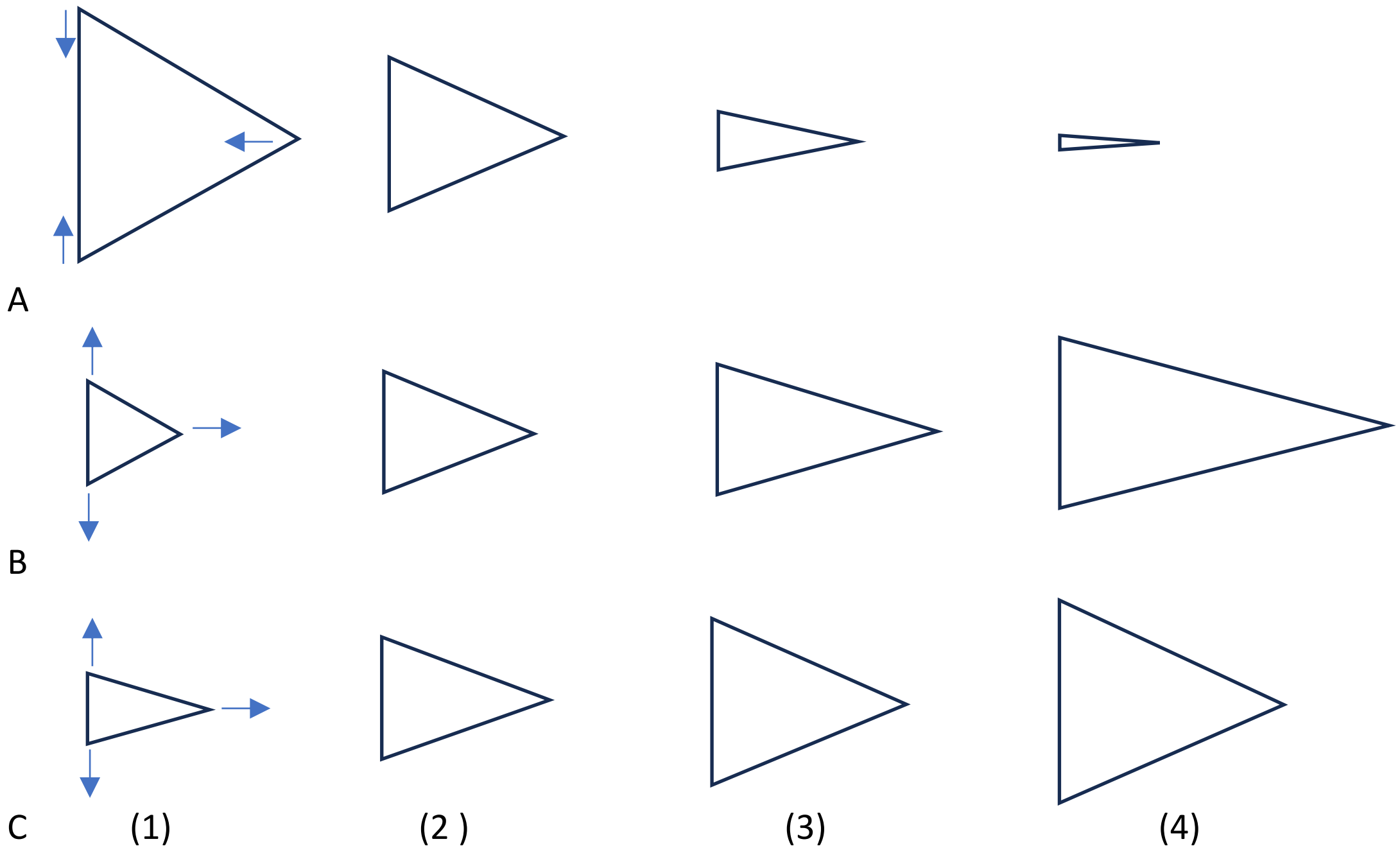

**Figure 10**. Qualitative description of the time evolution of the Lagrange triangle shape. Arrows indicate the direction of the bodies motion. **A**. The pure gravitational interaction is illustrated. The initial location of the bodies corresponds to that depicted in **Figure 2**. Gravity causes the deformation of the triangle, resulting in the eventual disappearing of one its sides corresponding to $CMS \to 1$. **B**. The time evolution of the Lagrange triangle under Coulomb interactions, corresponding the numerical experiments, illustrated with **Figures 5, 7** is depicted. CMS grows with time and comes to the saturation value. **C**. The time evolution of the Lagrange triangle under the Coulomb interactions, corresponding the numerical experiments, illustrated with **Figure 6** is shown. CMS is decreased with time and comes to the asymptotic value

**Conclusions**

We report the time evolution of the continuous measure symmetry calculated for the system of the three bodies interacting *via* the potential $(r)\sim\frac{1}{r}$. Gravitational and electrostatic interactions between the point bodies were addressed. The continuous measure of symmetry was calculated with the software Taylor Center. We conclude that the continuous measure of symmetry, seen as a dynamical variable, supplies a valuable qualitative information about the behavior of the three-body interacting systems. In the case of the pure gravitational interaction the three-body-system, deviated from its initial symmetrical configuration, described by the Lagrange equilateral triangle comes to collapse. This gravitational collapse is accompanied by the growth of the continuous measure of symmetry, which eventually attains its limiting value, namely $CMS \rightarrow 1$. When three point bodies interact *via* the Coulomb repulsive interaction, the time evolution of CMS is quite different. It should be emphasized, that for all of the studied initial configurations of the point charges and all of their charge-to-mass ratios, CMS always comes with time to its asymptotic value. Sometimes CMS grows in a monotonic way, sometimes is passes *via* the minimal value, but it always attains the asymptotic value, which evidences the stabilization of the shape of the triangle, constituted by the interacting bodies. This is an important and non-obvious qualitative conclusion emerging from our computer simulations. In our future studies we plan to address the temporal evolution of the continuous measure of symmetry calculated for the bodies interacting *via* a diversity of potentials.

**AKNOWLEDGEMENTS**

The authors are indebted to the creator of the software “Taylor Center” Dr. Alex Gofen for extremely useful discussions.